\documentclass[aps,pra,superscriptaddress,showpacs,twocolumn]{revtex4-2}
\usepackage{bm}
\usepackage{amsmath}
\usepackage{amssymb}
\usepackage{slashed}
\usepackage{float}
\allowdisplaybreaks
\usepackage{subcaption}
\usepackage{dcolumn}
\usepackage{graphicx}
\newcolumntype{.}{D{x}{}{-1}}

\newcolumntype{w}[1]{D{.}{.}{#1}}
\newcolumntype{L}{>{$}l<{$}}

\newcommand{\Za}{{Z\alpha}}

\usepackage{graphicx}
\usepackage{xcolor}
\usepackage{nicefrac}

\begin{document}

\title{Nuclear polarizability effects in $^3$He$^+$ hyperfine splitting}

\author{Vojt\v{e}ch Patk\'o\v{s}}
\affiliation{Faculty of Mathematics and Physics, Charles University,  Ke Karlovu 3, 121 16 Prague
2, Czech Republic}

\author{Vladimir A. Yerokhin}
\affiliation{Max–Planck–Institut f\"ur Kernphysik, Saupfercheckweg 1, 69117 Heidelberg, Germany}

\author{Krzysztof Pachucki}
\affiliation{Faculty of Physics, University of Warsaw,
             Pasteura 5, 02-093 Warsaw, Poland}

\begin{abstract}

The nuclear polarizability effects in hyperfine splitting  of light atomic systems
are not well known. The only system for which they were previously
calculated is the hydrogen atom, where these effects were shown to
contribute about 5\% of the total nuclear correction.
One generally expects the polarizability effects
to become more pronounced for composite nuclei. In the present work we
determine the nuclear polarizability correction to the hyperfine splitting in He$^+$
by comparing the effective Zemach radius deduced from the experimental
hyperfine splitting with the Zemach radius obtained from the electron scattering.
We obtain a surprising result that the nuclear polarizability of the helion
yields just 3\% of the total nuclear correction, which is smaller than for the proton.

\end{abstract}
\maketitle

\section{Introduction}

The hyperfine structure (HFS) in atoms and ions is determined not only by
the value of the nuclear magnetic moment but also by the distribution of the charge and
the magnetic moment over the nucleus and by the nuclear vector polarizability.
These effects cannot be calculated accurately at present and are the main
source of uncertainty in theoretical predictions.

The nuclear effects in HFS are typically divided into two parts: the elastic and the inelastic
ones. The elastic effects are expressed in terms of the charge and magnetic form factors,
whereas the dominant inelastic effect is the nuclear polarizability.
It is well known that the dominant nuclear effect is of the elastic kind and
is proportional to the so-called Zemach radius \cite{zemach}, which is the convolution of
the electric and the magnetic form factors.

Little is known about the nuclear polarizability in HFS, mainly due to the complexity
of its theoretical description. The effect is the most pronounced for
the muonic deuterium ($\mu$D) HFS, where it is supposed to be as large as the elastic nuclear
contribution. The theoretical predictions for the $\mu$D HFS \cite{kalinowski:18} are in
tension with the experimental results \cite{pohl:16}.

For the electronic atoms the nuclear polarizability effects are much smaller than for the muonic atoms
but are not negligible.
Even for hydrogen the inelastic effects were shown to be significant and yield about 5\% of
the elastic contribution \cite{tomalak:19}.
For other atomic systems the inelastic HFS contributions are merely unknown.
Low \cite{low:50} has derived a simple formula for the leading nuclear-structure contribution, treating the nucleus
as a composite system of protons and neutrons and avoiding the use of the elastic nuclear form factors.
Friar \cite{friar:05} estimated the inelastic contribution to HFS in deuterium beyond the Low formula and
concluded that it is not significant.
Khriplovich {\em et al} \cite{khriplovich:96, khriplovich:04} claimed to derive the leading logarithmic part
of the inelastic contribution, but later
it was demonstrated by one of us \cite{pachucki:07} that this contribution vanishes in a more complete treatment.
In that work a formula for the inelastic contribution to atomic HFS was derived, but its complexity
prevented any practical applications so far. So, the inelastic contribution to HFS in light atomic systems
is merely unknown at present.

In the absence of theoretical calculations of the nuclear polarizability correction, in
the present work we perform its determination from the experimental HFS splitting.
We rely on the fact that all HFS corrections originating from the
relativistic and quantum electrodynamics (QED) effects for
the point nucleus can be calculated up to very high accuracy
and that the elastic form factors of the nucleus can be extracted from
analyzing the electron scattering data.

We introduce the effective Zemach radius $\widetilde{r}_Z$ which includes the inelastic nuclear
contribution and can be accurately determined from high-precision experimental results for the HFS splitting.
On the other hand, the standard elastic Zemach radius ${r}_Z$ was determined from the electron-scattering
data by Sick \cite{sick:14}. The difference $\widetilde{r}_Z - r_Z$ gives us the result for the
nuclear polarizability correction.

The significance of the effective Zemach radius is that it can be used to obtain highly
accurate theoretical predictions for the HFS of the neutral helium, which will be a subject of
our forthcoming investigations.

\section{HFS in hydrogenic atoms}

The leading hyperfine splitting of an $1S$ state is delivered by the so-called Fermi energy $E_F$,
which is given by
\begin{equation}
E_F = \frac{8}{3}(Z\alpha)^4\,\frac{\mu^3}{m\,M}\,(1+k) \,,
\end{equation}
where  $Z$ and $M$ are the nuclear charge number and the mass, respectively,
$\mu$ is the reduced mass of the atom and
$k$ is the nuclear magnetic moment anomaly $k = (g-2)/2$, with the natural nuclear $g$ factor
defined as
\begin{align}
\vec\mu = \frac{Z\,e}{2\,M}\,g\,\vec I\,.
\end{align}
Here, $\vec\mu$ and $\vec I$ are the magnetic moment and the spin of the nucleus,
respectively. The natural $g$ factor is related to the standard nuclear $g_I$ factor by
\begin{align}
g=g_I\,\frac{M}{Z\,m_p}\,.
\end{align}
$g_I$ can be obtain from the recent measurement of the shielded $g_I$ in $^3$He$^+$ in Ref. \cite{blaum:22}
and the most accurate calculation of the shielding factor in Ref. \cite{pachucki:21:shielding}, namely
\begin{align}
g_I =&\ -4.255\,250\,699\,9\,(34)\,,
\end{align}
and therefore
\begin{align}
g =&\   -6.368\,307\,500\,5\,(51).
\end{align}

The complete hyperfine splitting is conveniently represented as
\begin{align}
E_\mathrm{hfs} =&\ E_F\,(1 + \delta)\,,
\end{align}
where $\delta$ represents the correction to the Fermi energy
due to relativistic, QED, and nuclear effects. Within
the approach of the nonrelativistic QED (NRQED),
$\delta$
is represented as
an expansion in terms of the fine-structure constant $\alpha$,
\begin{align} \label{eq:04}
\delta =&\  \kappa+ \delta^{(2)}+ \delta^{(3)} + \delta^{(4)} + \delta^{(1)}_\mathrm{nuc} + \delta^{(1)}_\mathrm{rec}+ \delta^{(2)}_\mathrm{nuc} + \delta^{(2)}_\mathrm{rec}\,,
\end{align}
where $\kappa $ the magnetic moment anomaly of the free electron, $\kappa = \alpha/(2\pi) + O(\alpha^2)$,
and $\delta^{(i)}$, $\delta^{(i)}_\mathrm{nuc}$, and $\delta^{(i)}_\mathrm{rec}$ are the
QED, nuclear, and recoil corrections of order $\alpha^i$, respectively.

The QED corrections of order
$\alpha^2$, $\alpha^3$, and $\alpha^4$ are given by
\cite{eides:01,tiesinga:21:codata18}
\begin{align}
\delta^{(2)}&\  =\frac{3}{2}\,(Z\,\alpha)^2 + \alpha\,(Z\,\alpha) \Bigl(\ln(2)-\frac{5}{2}\Bigr)\,,\\
\delta^{(3)}&\  = \frac{\alpha\,(Z\,\alpha)^2}{\pi}\!
\Big[\!-\frac{8}{3}\ln(Z\,\alpha)\Bigl(\ln(Z\,\alpha)-\ln(4)+\frac{281}{480}\Bigr)
\nonumber \\ &\
+ 17.122\,338\,751\,3-\frac{8}{15}\,\ln(2)+\frac{34}{225}\Bigr]
\nonumber \\ &\
+ \frac{\alpha^2\,(Z\,\alpha)}{\pi}\,0.770\,99(2) \,, \\
\delta^{(4)} &\  =\frac{17}{8}\,(Z\,\alpha)^4 +\alpha\,(Z\,\alpha)^3\,\Big[\Big(\frac{547}{48}-5\,\ln(2) \Big)\,\ln(Z\,\alpha)
\nonumber \\ &\
 -4.493\,23(3) +\frac{13}{24}\,\ln 2 + \frac{539}{288} \Bigr]
\nonumber
\\ &\
-\frac{\alpha^2\,(Z\,\alpha)^2}{\pi^2}\,\Bigl[\frac{4}{3}\,\ln^2(Z\,\alpha)+1.278\,\ln(Z\,\alpha) +10.0(2.5)\Bigr]
  \nonumber \\ & \
\pm  \frac{\alpha^3\,(Z\,\alpha)}{\pi^2}\,.
\end{align}
The last term in $\delta^{(4)}$ represents the estimate of the
unknown three-loop QED correction. We note that
the $\alpha(Z\,\alpha)^2$ part contains the improved numerical value from the Appendix A, whereas
the $\alpha(Z\,\alpha)^3$ numerical term includes higher orders in $Z\,\alpha$  for $Z=2$ from
Ref.~\cite{yerokhin:08:prl}.

\begin{table}
\caption{Contributions to HFS in ${}^3\mathrm{He^+}$ ion and determination of $\widetilde{r}_Z$.
The nuclear charge radius $r_C = 1.973\,(14)$~fm \cite{sick:14}.}
\begin{ruledtabular}
\begin{tabular}{lw{0.10}w{9.8}}
\multicolumn{1}{l}{Term} &
        \multicolumn{1}{c}{Value} &
            \multicolumn{1}{c}{$\times E_F$\,[kHz]}
\\
\hline\\[-5pt]
 $1$  & 1 &  -8\,656\,527.892\,(7)   \\
 $\kappa$  & 0.001\,159\,65 &   -10\,038.6  \\
 $\delta^{(2)}$ & 0.000\,127\,07   & -1\,100.0\\
 $\delta^{(3)}$ &-0.000\,019\,49   & 168.7\\
 $\delta^{(4)}$ &-0.000\,000\,75     & 6.5       \\
 $\delta_\mathrm{rec}^{(1)}$  &   -0.000\,012\,17\,(60) & 105.4\,(5.3)\\
 $\delta_\mathrm{nuc}^{(2+)}$  & -0.000\,002\,89(3) & 25.0\\
 $\delta_\mathrm{rec}^{(2)}$  &  -0.000\,001\,16\,(18)  & 10.1\,(1.6)\\ \\
 theory without  $\delta_\mathrm{nuc}^{(1)}$ & 1.001\,250\,26\,(63) &  -8\,667\,350.8\,(5.5)   \\
 experiment \cite{blaum:22}      & 1.001\,053\,77 & -8\,665\,649.865\,77\,(26)     \\
 $\delta_\mathrm{nuc}^{(1)}$  &    -0.000\,196\,49\,(63) & 1\,701.0\,(5.5)\\ \\
    $\widetilde{r}_Z$ this work  &   2.600\,(8)\,\mathrm{fm} \\
     $r_Z$ \cite{blaum:22}   &    2.608\,(24)\,\mathrm{fm} \\
    $r_Z$ \cite{sick:14} exp  &    2.528\,(16)\,\mathrm{fm}  \\
    $r_Z$ \cite{dinur:19} nucl. theo  &  2.539\,(3)(19)\,\mathrm{fm}  \\ \\
   $\widetilde{r}_Z - {r}_Z$(exp) =  & 0.072\,(18)\,\mathrm{fm}
\end{tabular}
\end{ruledtabular}
\end{table}

$\delta^{(1)}_\mathrm{nuc}$ is the leading $O(Z\,\alpha)$ nuclear structure correction.
It is a sum of the elastic contribution proportional to the Zemach radius $r_Z$
and the nuclear polarizability contribution. The Zemach radius is defined as
\begin{equation}
r_Z = \int d^3 r_1 \int d^3 r_2 \,\rho_E(\vec r_1)\,\rho_M(\vec r_2) \,|\vec r_1 - \vec r_2|\,,
\end{equation}
where $\rho_E$ and $\rho_M$ are the Fourier transforms of the electric and magnetic form factors
of the nucleus normalized to unity.

In this work we introduce the {\em effective} Zemach radius $\widetilde r_Z$ that includes
both the elastic and the inelastic contributions of order $\alpha$. It is related to
$\delta^{(1)}_\mathrm{nuc}$, by definition, as
\begin{align}\label{09}
\delta^{(1)}_\mathrm{nuc} =&\ -2\,Z\,\alpha\,\mu\,\widetilde r_Z\,.
\end{align}
The difference $\widetilde r_Z - r_Z$ can then be interpreted as the inelastic nuclear-polarizability
contribution.
In the present work, we determine $\delta^{(1)}_\mathrm{nuc}$ and, therefore, $\widetilde r_Z$
by taking the difference of the experimental HFS value and the theoretical prediction without
$\delta^{(1)}_\mathrm{nuc}$.

The higher-order nuclear-structure corrections and the recoil corrections are
much smaller than $\delta^{(1)}_\mathrm{nuc}$. They
will be calculated assuming that the distributions of the nuclear charge and
the magnetic moment are the same, $\rho_E(r) = \rho_M(r) \equiv \rho(r)$. 
In the present work, we will use the exponential and the Gaussian models 
for $\rho(r)$,
see Table~\ref{tab:models}. The parameters of the models will be fixed
by matching the Zemach radius to the experimental value
from Ref.~\cite{sick:14}. In order to estimate the model dependence of our
results, we take twice the difference of values obtained with the exponential
and the Gaussian models.

The $\alpha^2$ nuclear-structure correction is given by the sum of the relativistic and the
radiative corrections,
\begin{align}
\delta^{(2)}_\mathrm{nuc} = \delta^{(2)}_\mathrm{nuc,rel} + \delta^{(2)}_\mathrm{nuc,rad}\,.
\end{align}
The relativistic correction is given by \cite{kalinowski:18}
\begin{align}
\delta^{(2)}_\mathrm{nuc,rel}=&\ \frac{4}{3}\,(m\,r_C\,Z\,\alpha)^2 \\ & \times
\Big[-1+\gamma + \ln(2\,m\,r_{CC}\,Z\,\alpha) + \frac{r_M^2}{4\,r_C^2}\Big]\,, \nonumber
\end{align}
where
 $r_M$ is the root-mean-square magnetic radius
and $r_{CC}/r_C= 5.274\,565$ for the exponential charge distribution.
The numerical contribution of this correction is quite small,
$\delta^{(2)}_\mathrm{nuc,rel} = -5.4\times 10^{-8}$ for He$^+$.
Surprisingly, the next-order in $Z\,\alpha$ correction
yields a numerically larger contribution,
because it is approximately proportional to $m\,r_Z$,
instead of $(m\,r_C)^2$.
For this reason we replace $\delta^{(2)}_\mathrm{nuc,rel}$ by
$\delta^{(2+)}_\mathrm{nuc,rel}$, which is evaluated numerically in this work
for $Z = 2$. The resulting contribution is
\begin{align}
\delta^{(2+)}_\mathrm{nuc,rel} = -49.1(5)\times 10^{-8}\,.
\end{align}
The above value is obtained with the exponential model; its uncertainty represents the
expected model dependence and is obtained as twice the difference from the results 
obtained with the Gaussian model.

The radiative finite nuclear size correction is given within the exponential distribution
model by \cite{sgk:97}
\begin{align}
 \delta^{(2)}_\mathrm{nuc,rad}=&\  -2\,Z\,\alpha\,\mu\, r_Z
 \,\frac{\alpha}{\pi}\,\biggl( -\frac{5}{4} + \frac{2}{3}\,\ln\frac{\lambda^2}{m^2} - \frac{634}{315}\biggr)\,,
\end{align}
where the first terms comes from the electron self-energy and next two from the vacuum polarization.
Its numerical contribution for He$^+$ is
\begin{align}
 \delta^{(2)}_\mathrm{nuc,rad} =&\ -240.1(2.4) \times 10^{-8}\,,
\end{align}
assuming a similar 1\% uncertainty as for $\delta^{(2+)}_\mathrm{nuc,rel}$.

$\delta^{(1)}_\mathrm{rec}$ is the leading order (in $\alpha$) nuclear recoil correction,
given by \cite{pachucki:22}
\begin{align}
&\delta^{(1)}_\mathrm{rec} =  -\frac{Z\alpha}{\pi}\frac{m}{M}\frac{3}{8}\,\bigg\{g \bigg[\gamma - \frac74 + \ln(m\,r_{M^2})\bigg]\\
&- 4 \,\bigg[\gamma + \frac94 + \ln(m\,r_{EM})\bigg] - \frac{12}{g}\bigg[\gamma - \frac{17}{12} + \ln(m\, r_{E^2})\bigg]\bigg\}\,,\nonumber
\end{align}
where $\gamma \approx 0.577\,216$ is Euler's gamma constant,
\begin{equation}
\ln r_{EM} = \int d^3r_1 \int d^3 r_2\,\rho_E (\vec r_1)\,\rho_M(\vec r_2)\,\ln|\vec r_1 - \vec r_2|\,,
\end{equation}
and $\ln r_{M^2}$ and $\ln r_{E^2}$ defined analogously. Within
the exponential model,
\begin{align}
\ln m\,r_{EM} =\ln m\,r_{E^2} =\ln m\,r_{M^2} = -\ln \frac{\lambda}{m} -\gamma +\frac{23}{12}\,,
\end{align}
the numerical contribution for He$^+$ is
\begin{align}
\delta^{(1)}_\mathrm{rec} = -1\,217\,(60) \times 10^{-8}\,,
\end{align}
where we ascribed a 5\% uncertainty due to an approximate exponential parametrization of the
helion formfactors.

The higher-order recoil correction is the sum of the relativistic and the radiative-recoil contributions,
\begin{align}
\delta^{(2)}_\mathrm{rec} =&\ \delta^{(2)}_\mathrm{rec,rel} + \delta^{(2)}_\mathrm{rec,rad}\,.
\end{align}
The relativistic recoil correction was derived in Ref.~\cite{Bodwin:88}.
It has a finite point-nucleus limit and is given by
\begin{align}
\delta^{(2)}_\mathrm{rec,rel} =&\ (Z\,\alpha)^2\,\frac{\mu^2}{m\,M}\,\biggl\{-\biggl[1+7\,k + \frac{7}{1+k}\biggr]\,\frac{\ln(Z\,\alpha)}{4}
\nonumber \\ &\ \hspace{-10ex}
-\biggl[9 + 11\,k + \frac{23}{1+k}\biggr]\,\frac{\ln2}{4} + \frac{1}{36}\biggl[ -20 +31\,k+\frac{150}{1+k}\biggr]\biggr\}\,.
\end{align}
The numerical contribution for He$^+$ is $\delta^{(2)}_\mathrm{rec,rel} = -116.3\times 10^{-8}$.

The radiative recoil effect to HFS is in general unknown.
Karshenboim \cite{sgk:97} presented only a rough estimation for this correction in hydrogen.
Instead of using this estimate, we assume the logarithmic enhancement for this correction and estimate the uncertainty due to its omission  as
\begin{align}
\delta^{(2)}_\mathrm{rec,rad} =&\ \pm \delta^{(1)}_\mathrm{rec}\,\frac{\alpha}{\pi}\,\ln\frac{\lambda}{m}
 = \pm 18.4\times10^{-8}\,.
\end{align}

There are further higher-order corrections, such as the muonic and hadronic vacuum polarization, weak interactions, etc.
These corrections are smaller than the uncertainty of $\delta^{(1)}_\mathrm{rec}$ and thus neglected, see Ref. \cite{blaum:22} supplement.


\begin{table}
\caption{Various results for the exponential and Gaussian models of the nuclear charge and
magnetization distributions. $F_e$ is the charge distribution function, $V_C(r) = -\nicefrac{\Za}{r}\,F_e(r)$,
whereas $F_m$ is the magnetic moment distribution function, 
$H_{\mu} = \nicefrac{|e|}{4\pi}\, 
\bm{\alpha} \cdot 
\bm{\mu} \times \nicefrac{\bm{r}}{r^3}\, F_m(r)
$.
\label{tab:models}
}
\begin{ruledtabular}
\begin{tabular}{lcc}
\multicolumn{1}{l}{} &
        \multicolumn{1}{c}{Exponential} &
            \multicolumn{1}{c}{Gaussian}
\\
\hline\\[-5pt]
$\rho(q^2)$   &  $\frac{\lambda^4}{(\lambda^2+q^2)^2}$       &  $e^{-\nicefrac{a q^2}{2}} $\\[3pt]
$\rho(r)$     &  $\frac{\lambda^3}{8\pi}\, e^{-\lambda\,r}$  & $ \frac1{(2\pi a)^{3/2}}\, e^{-\nicefrac{r^2}{(2a)}}$ \\[3pt]
$r_C$         &  $\nicefrac{2\sqrt{3}}{\lambda} $                & $\sqrt{3 a}$ \\[3pt]
$r_Z$         &  $\nicefrac{35}{(8\,\lambda)}$                     & $4\,\sqrt{\nicefrac{a}{\pi}}$ \\[3pt]
$F_e(r)$      &  $1 - e^{-\lambda r}(1 + \nicefrac{\lambda r}{2}) $ 
                      & $ {\rm erf} \big( \frac{r}{\sqrt{2a}}\big)$ \\[4pt]
$F_m(r)$      &  $1 - e^{-\lambda r}\big( 1 + \lambda r +\nicefrac{(\lambda r)^2}{2}\big)$
                      & ${\rm erf}\big(\frac{r}{\sqrt{2a}} \big)
                         - \frac{\sqrt{2}\,r}{\sqrt{\pi a}} \, e^{-\nicefrac{r^2}{(2a)}}  $
\end{tabular}
\end{ruledtabular}
\end{table}

\section{Results  and Discussion}

In Table I we collect all theoretical contributions to the ground-state hyperfine splitting
of ${}^3\mathrm{He^+}$ ion.
It is interesting to note that
the recoil correction $\delta^{(1)}_\mathrm{rec}$ yields about 6\% of total nuclear contribution,
despite the smallness of the electron-nucleus mass ratio.

The difference of the theoretical prediction without
$\delta^{(1)}_\mathrm{nuc}$ and the experimental value from Ref.~\cite{blaum:22} determines
$\delta^{(1)}_\mathrm{nuc}$ and,
therefore, the effective Zemach radius.
The difference of the effective Zemach radius and the elastic Zemach radius obtained in Ref.~\cite{sick:14} from the
electron-scattering data yields  $\widetilde{r}_Z - r_Z = 0.072(18)$ fm.
We interpret this difference as the effect of the nuclear polarizability. It is remarkable that
its numerical contribution is quite small, just about 3\% of the elastic nuclear contribution.
This is even smaller than the inelastic contribution for hydrogen HFS of 5\% \cite{tomalak:19}.
We find this small result very surprising. At present we do not have an explanation why the inelastic
nuclear contribution for a helion as a composite nucleus is so smaller than for the proton.
It should be noted that the authors of Ref.~\cite{blaum:22}
claimed to evaluate the nuclear polarizability correction and obtained a very small result
(vanishing within their uncertainty), so within their uncertainty $\widetilde{r}_Z = r_Z$.

Summarizing and bearing in mind the discrepancy for $\mu$D HFS \cite{kalinowski:18} and for $^6$Li HFS \cite{sun:23},
there is no yet any comprehensive theory for the inelastic (polarizability) correction to the atomic HFS with the composite nucleus.

\begin{acknowledgments}
Valuable communications from Dr. Bastian Sikora are gratefully acknowledged.
K.P. and V.P. acknowledge support from the National Science Center (Poland) Grant No. 2017/27/B/ST2/02459.
\end{acknowledgments}

\appendix

\section{One-loop $\alpha\,(Z\,\alpha)^2$ self-energy correction to HFS}

In the previous work \cite{pachucki:96} devoted to $\alpha\,(Z\,\alpha)^2$ one-loop self-energy contribution the hyperfine splitting
bugs crept into formulas for intermediate contributions, while the final result was correct.
Here, we remove all these bugs and present numerical integrals with the higher precision, what might be relevant in
future studies of HFS is light atomic systems.

Using the notation from the former work \cite{pachucki:96} and thus setting for convenience $Z=1$, the one-loop self energy contribution
to HFS in hydrogen-like system is represented in terms of dimension-less function $F$.
\begin{align}
\Delta E = E_F\,\frac{\alpha}{\pi}\,\alpha^2\, F\,,
\end{align}
which is split into three parts
\begin{align}
F = F_L + F_M + F_H \label{a2}
\end{align}

The low energy part
\begin{align}
F_{L} =&\  {{781}\over {18}} + {{4\,{{\pi }^2}}\over {3}}  -
{{166\,\ln (2)}\over 3} - {{2\,{{\ln (2)}^2}}\over 3} - 4\,\ln (\alpha)
\nonumber \\&\
+ 8\,\ln (2)\,\ln (\alpha) -
 {{8\,{{\ln (\alpha)}^2}}\over 3}
+ 2\,\ln (\epsilon) - 4\,\ln (2)\,\ln (\epsilon)
\nonumber \\ &\
+ {{8\,\ln (\alpha)\,\ln (\epsilon)}\over 3} -
   {{2\,{{\ln (\epsilon)}^2}}\over 3} +n_1+n_2\,,
\end{align}
is expressed in terms of two integrals Eqs. (26,27) of Ref. \cite{pachucki:96}
\begin{align}
n_1 =&\ -0.085\,740\,323\,701\,4\,, \\
n_2 =&\ \hspace{3ex}  0.067\,496\,936\,500\,3 \,,
\end{align}
which we present here with much higher precision.

The middle energy part
\begin{align}
F_M =&\ F_{M1}+F_{M2}+F_{M3} + F_{M4}
\end{align}
consists of four subparts
\begin{align}
F_{M1} =&\ \frac{1}{2}\left[1-\frac{1}{2}\,\ln\left(\frac{2\,\alpha}{\rho}\right)\right]\,,\\
F_{M2} =&\ -\frac{8}{3}\,\left[ \frac{1}{2}- \ln\Bigl(\frac{2\,\alpha}{\rho}\Bigr)\right]\,\ln\Bigl(\frac{m}{\mu}\Bigr)\,,\\
F_{M3} =&\ 4\,\ln\Bigl(\frac{m}{\mu}\Bigr)-\frac{1}{2}\,,\\
F_{M4} =& \ln\Bigl(\frac{2\,\alpha}{\rho}\Bigr)\,.
\end{align}
Their sum is
\begin{align}
F_M =&\
{{\frac{8}{3}\,\ln \Bigl({m\over {\mu }}\Bigr)}} +
   {{\frac{3}{4}\,\ln \Bigl({{2\,\alpha}\over {\rho }}\Bigr)}} +
   {{\frac{8}{3}\,\ln \Bigl({m\over {\mu }}\Bigr)
\,\ln \Bigl({{2\,\alpha}\over {\rho }}\Bigr)}} \\
=&\
\frac{20}{9} - \frac{8}{3}\,\ln(2\,\epsilon) + \frac{107}{36}\,\ln\Bigl(\frac{2\,\alpha}{\rho}\Bigr)
- \frac{8}{3}\,\ln(2\,\epsilon)\,\ln\Big(\frac{2\,\alpha}{\rho}\Big)\label{FM}
\end{align}

The high energy part is
\begin{align}\label{FH}
F_H =&\
  -{{335}\over {36}} - {{11\,{{\pi }^2}}\over 18} + {{190\,\ln (2)}\over 9} +
   {{2\,{{\ln (2)}^2}}\over 3} + {{2\,\ln (\epsilon)}\over 3}
   \nonumber \\ &\
   + {{20\,\ln (2)\,\ln (\epsilon)}\over 3} +
   {{2\,{{\ln (\epsilon)}^2}}\over 3}
   + {{107\,\ln (\rho )}\over {36}}
   \nonumber \\ &\
    - {{8\,\ln (2)\,\ln (\rho )}\over 3} -
   {{8\,\ln (\epsilon)\,\ln (\rho )}\over 3} - {{5\,{\zeta}(3)}\over 4}\,.
\end{align}

The final result is a sum of three parts as given in Eq. (\ref{a2})
\begin{align}
F  =&\  n_1+n_2+
{{1307}\over {36}} + {{13\,{{\pi }^2}}\over {18}}
 - {{407\,\ln (2)}\over {12}} - {{8\,{{\ln (2)}^2}}\over 3}
 \nonumber \\ &\
 - {{37\,\ln (\alpha)}\over {36}}
+ {{16\,\ln (2)\,\ln (\alpha)}\over 3} -
   {{8\,{{\ln (\alpha)}^2}}\over 3} - {{5\,{\zeta}(3)}\over 4}\,,
\end{align}
and the constant term equals to $ 17.122\,338\,751\,3$\ldots

In the calculation of individual parts, we used sometimes the photon mass $\mu$  regularization
and equivalently, the photon cut-off $\epsilon$. The conversion formulas  from $\epsilon$ to $\mu$ are
the following
\begin{align}
\ln(\epsilon) =&\ \ln\Big(\frac{\mu}{2}\Big) + \frac{5}{6}\,,\\
\ln^2(\epsilon) =&\ \bigg[\ln\Big(\frac{\mu}{2}\Big) + \frac{5}{6}\bigg]^2 + \frac{31}{36} -\frac{\pi^2}{12}\,,
\end{align}

\end{document}